\documentclass{elsart}
\usepackage{amssymb,amscd,amsmath,graphicx,cite}
\begin{document}
\runauthor{Goychuk, H\"anggi}
\begin{frontmatter}

\title{Quantum Two-State Dynamics Driven by Stationary
Non-Markovian Discrete Noise: Exact Results}
\author{Igor Goychuk\corauthref{cor1}},
\author{ Peter H\"anggi}


\address{Universit\"at Augsburg, Institut f\"ur Physik,
 Universit\"atsstr. 1, D-86135 Augsburg, Germany}

\corauth[cor1]{Corresponding author,  
e-mail: goychuk@physik.uni-augsburg.de }

\begin{abstract}
We consider the problem of stochastic averaging of a 
quantum two-state dynamics
driven by non-Markovian, discrete noises of the continuous time
random walk type (multistate renewal processes). The emphasis is put
on the proper averaging over the stationary noise realizations
corresponding, e.g.,  to a stationary environment. A two state
non-Markovian process with an arbitrary non-exponential distribution
of residence times (RTDs) in its states with a finite mean residence
time provides a  paradigm. For the case of a two-state quantum
relaxation caused by such a classical stochastic field we obtain the
explicit exact, analytical expression for the averaged
Laplace-transformed relaxation dynamics. In the limit of Markovian
noise (implying an exponential RTD), 
all previously known results are recovered.
We exemplify new more general results for the case of
non-Markovian noise with a biexponential RTD. The averaged,
real-time relaxation dynamics is obtained in this case by 
numerically exact solving of a resulting algebraic polynomial problem. 
Moreover, the case of manifest non-Markovian noise with an infinite
range of temporal autocorrelation (which in principle is not
accessible to any kind of perturbative treatment) is studied, both
analytically (asymptotic long-time dynamics) and numerically (by a
precise numerical inversion of the Laplace-transformed averaged
quantum relaxation).

\end{abstract}


\end{frontmatter}

\section{\label{Intro}Introduction}

The influence of a stochastic environment on relaxation and
charge transfer processes
in condensed media \cite{Marcus,Hush,Levich} is a longstanding 
problem of prime importance
in chemical and statistical physics 
\cite{Hanggi,Grifoni,Kohler,Jortner,May,Petrov,review}. 
In this context,  exactly solvable models  are
rather rare. The method of a stochastic Hamiltonian by Anderson and Kubo
\cite{Anderson53,Kubo54,Kubo62}, known also under the label of stochastic Liouville
equation (SLE) approach\cite{Kubo63,Kubo69,Burshtein,
Haken72,Haken73,BlumenSilbey,Fox,MukamelOppenRoss,ReinekerBook} presents a common methodology
that has been employed over decades \cite{ReinekerBook,LindWest,review}:
It is based on the simplifying representation of a stochastic environment
by a   classical stochastic field that acts on the quantum system
of interest without a  feedback mechanism. This set up thus necessarily inherits some
shortcomings. In particular, quantum states become asymptotically
populated with equal weight as if the environment temperature were infinite. This
shortcoming can be cured in an {\it ad hoc}  manner by adding an extra term to the
stochastically averaged Liouville equation to ensure a proper thermal
equilibrium. What  ``proper'' means, however, is  specific to the problem under consideration.
Therefore, the mentioned shortcoming can be overcome rigorously  only within a full
quantum-mechanical treatment of the total, coupled system-environment 
dynamics \cite{Stockburger,Shao}.
Notwithstanding this  principal difficulty, the SLE approach remains useful
and popular over the years. The method is particularly appealing
because it allows to extract  exact
results for a number of interesting physical models \cite{Haken72,OvchinErikh,Fox,ReinekerBook,
LindWest,Kayanuma84,Kayanuma85,ShaoZerbeHan98}.

In view of the  central limit theorem, the use of
classical random forces with a Gaussian statistics is most frequently employed
for this sort of semiclassical modeling. The  case of Gaussian  white noise
serves frequently as a simple model for the corresponding
classical stochastic bath; it formally corresponds to a heat
bath with an infinite spectrum of excitations.
For this noise model there exists a number of  exact results
\cite{Haken72,OvchinErikh,Fox,ReinekerBook,
LindWest,ShaoZerbeHan98}.
Realistic  thermal baths possess, however, a spectrum with  cut-off at finite energies.
This circumstance then renders temporal autocorrelations that decay on a finite time scale.
The use of stationary
Gaussian Markov noise with an exponentially decaying temporal
autocorrelations (Ornstein-Uhlenbeck process) presents one of the simplest models for the corresponding
colored noise \cite{Fox,HanggiJung}. We remark, however, that even for the archetype case  of a
two-state tunneling system this colored noise model  cannot be solved exactly, except
 for some special situations, see, e.g., in Ref. \cite{Kayanuma84} for
a stochastic Landau-Zener model. In practice, one is forced to invoke additional approximations;
such as in the case of a weakly
colored noise the method of  cumulant expansions
\cite{MukamelOppenRoss,Fox,KampenBook}. It is equivalent to an expansion in a small parameter
(the Kubo number) which is the root mean
square (rms) of the fluctuations of the
characteristic coupling energy (in  frequency units)
times the autocorrelation time of corresponding bath fluctuations \cite{KampenBook}.
Other perturbation schemes can also be used \cite{HanggiJung}.
In particular, the opposite limit of large Kubo numbers (quasi-static
fluctuations) also allows for a consistent perturbative treatment
\cite{KampenBook,HanggiJung,Schulten}.

Nonperturbative approaches are, however, much more appealing.
Continuous state fluctuations can be approximated with jump-like stochastic
processes possessing a large
number of discrete states. An example presents a  discretization of the spatial degree of freedom
of a continuous diffusion
process. The class of Markovian
discrete state noises provide  a rather general
framework for a formally
exact averaging without using any kind of perturbation
expansion \cite{Kubo62,Burshtein,FrischBrissaud1,FrischBrissaud2,Hanggi78,
Hanggi80,Hanggi81}.
The stochastic
path integral approach \cite{Wiener,FeynmannBook}, adapted for such jump
processes \cite{Burshtein,FrischBrissaud1}
is particularly convenient. There exist yet other powerful approaches, cf. in the Refs.
\cite{FrischBrissaud2,ShapLog1,Klyatskin,ShapLog,HanggiProceedings}.
The two-state Markovian noise (or dichotomous noise) presents
the archetype discrete state process which allows for an exact study of
stochastically driven  two-level quantum systems
\cite{AverPouq,ShapLog1,Reineker1,Reineker2,PRE95_3,PLA96,Iwan,AnkerPech}.
Furthermore, the multistate case of an exciton transfer
in molecular
aggregates with many independent noise sources modeled by
independent two-state Markovian noises
is also solvable, in the sense that it can be reduced
to solving numerically an algebraic relation
\cite{Reineker3}. This
dichotomous noise can serve to model a
pseudo-spin 1/2 stochastic bath degree of freedom.
In the case of electron transfer in molecular systems such pseudo-spin
stochastic variables can simulate, for example, the bistable
fluctuations of a charged
molecular group nearby the donor, or the acceptor site, or also the conformational
fluctuations of a bistable molecular bridge \cite{JCP95}.

The method of a stochastic path averaging can readily be generalized
onto non-Markovian jump processes  of
the renewal (or continuous time random walk)
type \cite{Kampen79,Bursh86,Chvosta}.
Such processes are then characterized by the set of
residence time distributions (RTDs) $\psi_j(\tau)$ in the noise states $j$
and the probabilities $p_{ij}$ for undergoing a transition from state $j\to i$
at the end of each residence time interval (RTI) into another noise
state $i$.
The lengths of subsequent
 RTIs are mutually independent. \footnote{This does not necessarily imply  that the
 process is  Markovian: If the
  RTD of some state is non-exponential
 the process is clearly non-Markovian and exhibits memory effects.
 This can be  understood due to the following reasoning which can
 be traced back to Ref. \cite{Feller}.
 For a memoryless, Markovian  process the survival
 probability $\Phi_j(\tau):=\int_{\tau}^{\infty}\psi_j(\tau')d\tau'$
 in any state $j$
 must obviously assume the following property:
 $\Phi_j(\tau_1+\tau_2)=\Phi_j(\tau_1)
 \Phi_j(\tau_2)$. The only nontrivial solution of this latter functional
 equation reads
 $\Phi_j(\tau)=\exp(-\gamma_j\tau)$, where $\gamma_j>0$ is
 the rate to leave the state $j$. }
 This constitutes the crucial ingredient that allows for an exact averaging for such non-Markovian
 processes \cite{Kampen79}.
 Attention must be paid, however, to the  stationarity of the process, i.e.
to the proper averaging over the stationary noise realizations in a stationary
environment. This issue
is not trivial and requires special consideration:
Such a stationary averaging is  possible only if the average residence times
are finite for all discrete noise states.  The stationary noise averaging
can be performed exactly in the Laplace-domain for any non-Markovian process of the
considered type and arbitrary quantum dynamics \cite{Bursh86,Goychuk04}.
In particular, the Laplace-transform of the corresponding quantum propagator
can be written in a general analytical form \cite{Goychuk04}.
It yields a complex expression which practically can be elaborated on explicitly
in some special cases only. In this work, we present  analytical
expressions for averaged quantum two-state dynamics driven by a
symmetric two-state
non-Markovian noise with arbitrary RTD possessing a finite mean time.
Our results generalize the prior results of Refs.
\cite{AverPouq,ShapLog1,AnkerPech} and reduce to the latter ones in
the  particular case of Markovian noise.

The  paper is structured as follows. We first review the
general approach and present the main result for the
Laplace-transformed stationary-averaged quantum propagator in Sec. 2.
A specific application of this general result to the Kubo oscillator is given
in Sec. 3.
Explicit results for a quantum two-state dynamics driven by a two-state
{\it non-Markovian noise} are provided in Sec. 4.
The limit of Markovian noise and a special
non-Markovian case are studied for the two-state quantum dynamics
in Sec. 5. The case of manifest non-Markovian noise with infinite mean
autocorrelation time is considered in Sec. 6.
In Sec. 7 we provide a resume of our results.

\section{Stochastic Liouville-von-Neumann equation and corresponding averaged quantum
propagator}

To start out, let us consider an arbitrary quantum system with
a Hamilton operator $\hat H[\xi(t)]$ which depends on a classical,
 noisy parameter
 $\xi(t)$. This stochastic process $\xi(t)$ can take on either
continuous or discrete values.  Accordingly, the Hamiltonian
$\hat H$  randomly in time  acquires different operator values
 $\hat H[\xi(t)]$ which generally do {\it not} commute, i.e.,
$[\hat H[\xi(t)],\hat H[\xi(t')]]\neq 0$.

The prime objective  is then to noise-average the corresponding quantum dynamics
which is characterized by the Liouville-von-Neumann equation
\begin{equation}\label{Liouville}
\frac{d}{dt} \rho (t)= -i \mathcal{ L}[\xi(t)]\rho(t) \;,
\end{equation}
for the density operator $\rho(t)$ over the realizations
of the noise $\xi(t)$. $\mathcal{ L}[\xi(t)]$
in Eq. (\ref{Liouville})
denotes the quantum Liouville
superoperator, $\mathcal{ L}[\xi(t)](\cdot)=\frac{1}{\hbar}
[\hat H[\xi(t)], (\cdot)]$. In other words, the posed challenge consists in evaluating
the noise-averaged propagator
\begin{equation}\label{propagator}
\langle S (t_0+t,t_0) \rangle =
\langle \mathcal{ T} \exp[-i\int_{t_0}^{t_0+t}
\mathcal{ L}[\xi(\tau)] d\tau] \rangle,
\end{equation}
where $\mathcal{ T}$ denotes the time-ordering operator.

\subsection{Non-Markovian {\it vs.} Markovian discrete state
fluctuations}

We next specify this task for noise assuming $N$ discrete  states
  $\{\xi_i\}$.
The noise is generally assumed to present a non-Markovian renewal process which is
fully characterized by
the set of  transition probability densities $\psi_{ij}(\tau)$
for making random transitions within the time
interval $[\tau,\tau+d\tau]$ from
the  state $j$ to the state
$i$. These probability densities are necessarily  positive and  obey the
normalization conditions
\begin{equation}\label{norm}
\sum_{i=1}^{N}\int_{0}^{\infty}
\psi_{ij}(\tau)d\tau =1 \;,
\end{equation}
for all $j=1,2,...,N$.

The subsequent residence time-intervals between 
jumps are assumed to be mutually
uncorrelated.
The residence time distribution (RTD) $\psi_j(\tau)$ in the state $j$
reads
\begin{equation}\label{RTD}
\psi_j(\tau)=\sum_i \psi_{ij}(\tau)=-\frac{d\Phi_j(\tau)}{dt}.
\end{equation}
 The survival probability $\Phi_j(\tau)$
of the state $j$ follows then as
\begin{equation}\label{survival}
\Phi_j(\tau)=\int_{\tau}^{\infty}
\psi_j(\tau)d\tau.
\end{equation}
This constitutes the  general scheme for continuous time
random walk
(CTRW) theory \cite{MontrolWeiss,LaxSher,Shlesinger,Hughes}.

Several particular descriptions
used for such non-Markovian processes of the renewal type
are worth mentioning. The approach in Ref. \cite{Kampen79}
in terms of time-dependent aging rates $k_{ij}(t)$ for the transitions
from state $j$ to state $i$ corresponds to a particular choice, reading
\begin{equation}\label{kamp}
\psi_{ij}(\tau):=k_{ij}(\tau)\exp[-\sum_i\int_{0}^{\tau}k_{ij}(t)dt].
\end{equation}
The {\it Markovian} case corresponds to time-independent transition rates
$k_{ij}(\tau)=const$.
Any deviation of $\psi_{ij}(\tau)$ from this Markovian case then
in turn yields a
time-dependence of the transition rates $k_{ij}(\tau)$ which
amounts to a non-Markovian behavior. Furthermore,
the survival probability $\Phi_j(\tau)$ in the
state $j$  is determined by
\begin{equation}\label{kamp2}
\Phi_j(\tau)=\exp[-\sum_{i=1}^{N}
\int_{0}^{\tau}k_{ij}(t)dt]
\end{equation}
and Eq. (\ref{kamp})  can be recast as
\begin{equation}\label{kamp3}
\psi_{ij}(\tau):=k_{ij}(\tau)\Phi_j(\tau).
\end{equation}
The introduction of such time-dependent ``aging''
rates presents one possibility to describe  non-Markovian memory effects;
it is not unique though.
A different
scheme  follows by defining  \cite{Chvosta}:
\begin{equation}\label{alter}
\psi_{ij}(\tau):=p_{ij}(\tau)\psi_j(\tau)
\end{equation}
 with $\sum_i p_{ij}(\tau)=1$.
The physical interpretation is as follows: The process remains
in a state $j$ for a random time interval characterized by the
probability density $\psi_j(\tau)$. At the end of this time interval
it jumps into another state $i$ with a generally
time-dependent conditional probability $p_{ij}(\tau)$.
Such an interpretation is frequently used in the continuous time random
walk theory. Evidently,
any process of the considered type can be interpreted in this way.
By equating Eq. (\ref{kamp3}) and Eq. (\ref{alter}) and taking into
account that $\psi_j(\tau):=-d\Phi_j(\tau)/d\tau$ one  deduces
that the approach in Ref. \cite{Kampen79} can be related to that
in Ref. \cite{Chvosta}
with the time-dependent transition probabilities
\begin{equation}\label{corres}
p_{ij}(\tau)=\frac{k_{ij}(\tau)}{\sum_i k_{ij}(\tau)}
\end{equation}
and with the non-exponential probability densities $\psi_j(\tau)$, \\
i.e.,
$\psi_j(\tau)= \gamma_j(\tau)\exp[-\int_{0}^{\tau}\gamma_j(t)dt]$ with
$\gamma_j(\tau):=\sum_ik_{ij}(\tau)$.

The description of non-Markovian effects with the
time-dependent transition probabilities $p_{ij}(\tau)$,
is rather difficult
to derive from the sample trajectories of
an experimentally {\it observed} random process $\xi(t)$.
The same holds true for
the concept of time-dependent rates.
These rates cannot be measured directly from
the set of stochastic sample trajectories.
On the contrary, the RTD $\psi_j(\tau)$ and the
{\it time-independent} $p_{ij}$ (with $p_{ii}:=0$)
can routinely be deduced from measured
sample trajectories, say, in a {\it single-molecular} experiment.
 The study of the
statistics of the residence time-intervals allows one to obtain
the corresponding probability densities $\psi_j(\tau)$
and, hence, the survival probabilities $\Phi_j(\tau)$. Furthermore,
the statistics of
the transitions from one state into all other states allows one to derive
the corresponding conditional probabilities $p_{ij}$. From this primary
information
a complementary interpretation of experimental data in terms of
time-dependent rates
$k_{ij}(\tau)$ can readily be given as
\begin{equation}\label{reduce}
k_{ij}(\tau)=-p_{ij}\frac{d\ln[\Phi_j(\tau)]}{d\tau},
\end{equation}
if one prefers this particular ``language'' to describe the
non-Markovian effects.
Moreover, the description with a constant set
$p_{ij}$ provides a consistent approach to describe  stationary
realizations of $\xi(t)$, and consequently to  find the corresponding  averaged quantum evolution
 \cite{Goychuk04}. It is this reasoning that we shall follow in the following.

\subsection{Averaging the quantum propagator}

The task of performing the noise-averaging of the quantum dynamics
in Eq. (\ref{propagator}) can be solved exactly because we can make use of the piecewise
constant character of the noise realizations $\xi(t)$
\cite{Burshtein,FrischBrissaud1,Kampen79}.
Indeed, let us
consider the time-interval $[t_0,t]$ and let us take a frozen realization
of $\xi(t)$ assuming $k$ switching events within
this time-interval at the time-instants $t_i$,
\begin{equation}\label{time}
t_0<t_1<t_2<...<t_k<t.
\end{equation}
Correspondingly, the noise takes on the values
$\xi_{j_0},\xi_{j_1},...,\xi_{j_k}$ in the time sequel.
Then, the propagator $S(t,t_0)$ reads
\begin{equation}\label{pathU}
S(t,t_0)=e^{-i \mathcal{ L}[\xi_{j_k}](t-t_{k})}
e^{-i \mathcal{ L}[\xi_{j_{k-1}}](t_k-t_{k-1})}...e^{-i \mathcal{ L}
[\xi_{j_0}](t_1-t_{0})} \;.
\end{equation}
Let us further assume  that the process $\xi(t)$ has been {\it prepared}
in the state $j_0$ at $t_0$. Then,
the corresponding $k-$times probability density
for this noise realization reads
\begin{equation}\label{prob}
P_k(\xi_{j_k},t_{k};\xi_{j_{k-1}},t_{k-1};...;\xi_{j_1},t_{1}|\xi_{j_0},t_{0})=
\Phi_{j_k}(t-t_k)\psi_{j_kj_{k-1}}(t_k-t_{k-1})...
\psi_{j_1j_0}(t_1-t_0)
\end{equation}
for $k\neq 0$ and $P_0(\xi_{j_0},t_0)=\Phi_{j_0}(t-t_0)$ for $k=0$.
In order to obtain the noise-averaged propagator
$\langle S(t|t_0,j_0)\rangle$
conditioned on such a {\it nonstationary} initial noise preparation
in the state $j_0$
one has to
average (\ref{pathU}) with the probability measure in (\ref{prob})
(for $k=\overline{0,\infty}$).
This task can be readily be performed  by use of the
Laplace-transform method [it will be denoted in the following as $\tilde A(s):=
\int_{0}^{\infty}\exp(-s\tau) A(\tau) d\tau$ for any time-dependent
quantity $A(\tau)$]. The result for
$\langle \tilde S(s|t_0,j_0)\rangle=
\int_{0}^{\infty}\exp(-s\tau) \langle S(t_0+\tau|t_0,j_0)
\rangle d\tau$ thus reads \cite{Kampen79,Goychuk04}
\begin{equation}\label{result1}
\langle \tilde S(s|t_0,j_0)\rangle=\sum_i \Big
(\tilde A(s)[I-\tilde B(s)]^{-1} \Big )_{i j_{0}},
\end{equation}
where the matrix-superoperators  $\tilde A(s)$ and
$\tilde B(s)$ read in components
\begin{equation}\label{aux1}
\tilde A_{kl}(s):=\delta_{kl}
\int_{0}^{\infty}\Phi_l(\tau)e^{-(s+i \mathcal{ L}[\xi_{l}])\tau}
d\tau,
\end{equation}
and
\begin{equation}\label{aux2}
\tilde B_{kl}(s):=
\int_{0}^{\infty}\psi_{kl}(\tau)e^{-(s+i \mathcal{ L}[\xi_{l}])\tau}
d\tau\;,
\end{equation}
correspondingly, and $I$ denotes the unit matrix.

To obtain the stationary noise average of the propagator it necessary to average
(\ref{result1}) over the stationary initial probabilities $p_{j_0}^{st}$.
The averaging over the initial distribution alone
is, however, not sufficient to arrive at the stationary noise-averaging
in the case of non-Markovian processes since the noise realizations
constructed in the way still are generally  not stationary. This principal
problem is rooted in the
following observation:
By preparing the quantum system at $t_0=0$ in a nonequilibrium
state $\rho(0)$, the noise is taken
 at random in some initial state
 $\xi_{j_0}$ with the probability $p_{j_0}^{st}$ (stationary noise).
 However, every time when we  repeat the preparation of the quantum
 system in its initial state, the noise will
 already occupy a (random) state $\xi_{j_0}$
 for some unknown, random time interval $\tau_{j_0}^{*}$ (setting a clock at
 $t_0=0$ sets the initial time for the quantum
 system, but not for the noise, which is assumed to start in the infinite past).
Therefore, in a stationary
setting a proper averaging over this unknown
time $\tau_{j_0}^{*}$
is necessary. The corresponding procedure
 implies that the mean residence time $\langle\tau_j\rangle$ is
finite, $\langle\tau_j\rangle\neq \infty$, and
yields a different residence time
distribution for the initial noise state, $\psi_j^{(0)}(\tau)$; 
namely, it is evaluated to read
$\psi_j^{(0)}(\tau)=\Phi_j(\tau)/\langle \tau_j\rangle$ \cite{cox}.
Only for Markovian processes where  $\Phi_j(\tau)$ is strictly
exponential,
does $\psi_j^{(0)}(\tau)$ coincide
with $\psi_j(\tau)$.
Using this $\psi_j^{(0)}(\tau)$ instead of $\psi_j(\tau)$
for the first sojourn in the corresponding state
and for the {\it time-independent} set $p_{ij}$,
the noise realizations
become stationary \cite{cox,Goychuk04,Bursh86}. The corresponding expression
for the quantum propagator averaged over such stationary noise realizations
has been derived in Ref. \cite{Goychuk04}, cf. Eqs. (25), (29) therein.
In a slightly  more general form it reads
\begin{equation}\label{final}
\langle \tilde S(s)\rangle=\langle \tilde S(s)\rangle_{static}
-\sum_{ij} \Big
(\tilde C(s)-\tilde A(s)[I-P\tilde D(s)]^{-1} P\tilde A(s)
\Big )_{i j}\frac{p_j^{st}}{\langle \tau_j\rangle},
\end{equation}
where $\langle \tilde S(s)\rangle_{static}$ is the Laplace-transform of
the statically averaged Liouville propagator
\begin{equation}\label{static}
\langle S(\tau)\rangle_{static}:=\sum_k e^{-i \mathcal{ L}[\xi_{k}]\tau}p_k^{st},
\end{equation}
$p_j^{st}=\lim_{t\to\infty}p_j(t)$ are the stationary probabilities
of noise states which are determined by a
system of linear algebraic equations \cite{Goychuk04,Bursh86},
\begin{eqnarray}\label{eqpop}
\frac{p_j^{st}}{\langle \tau_j\rangle}=\sum_n p_{jn}\frac{p_n^{st}}
{\langle \tau_n\rangle}\;,
\end{eqnarray}
and $P$ is the matrix of transition
probabilities $p_{ij}$, i.e. the ``scattering matrix'' of the random process
$\xi(t)$. Furthermore, the auxiliary matrix operators $\tilde { C}(s)$
and $\tilde { D}(s)$ in (\ref{final}) read in components:
\begin{equation}\label{aux1new}
\tilde C_{kl}(s):=\delta_{kl}
\int_{0}^{\infty}e^{-(s+i \mathcal{ L}[\xi_{l}])\tau}
 \int_{0}^{\tau}\Phi_l(\tau')d\tau'd\tau
\end{equation}
and
\begin{equation}\label{aux2new}
\tilde D_{kl}(s):=\delta_{kl}
\int_{0}^{\infty}\psi_l(\tau)e^{-(s+i \mathcal{ L}[\xi_{l}])\tau}
 d\tau \;.
\end{equation}

We remark here that this very same averaging procedure can be applied to any system of
linear stochastic differential equations.

\section{\label{KuboOsc}An archetype case: The Kubo oscillator}

A simple but very instructive application of this general procedure is the noise-averaging of
the Kubo oscillator \cite{Kubo62,Shlesinger}; reading
\begin{equation}\label{kubo-osc}
\dot x(t)=i\epsilon[\xi(t)] x(t)\;.
\end{equation}
This particular equation emerges in various situations such as in the theory of optical line shapes,
 nuclear magnetic resonance
\cite{Kubo62,Anderson53}, and also for
single molecule
spectroscopy \cite{JungSilbey}. In the context of the stochastic theory
of spectral line shapes \cite{Kubo62,Anderson53,JungSilbey},
$\epsilon[\xi(t)]$
in Eq. (\ref{kubo-osc}) corresponds to a stochastically modulated
frequency of quantum
transitions between the levels of a ``two-state atom'', or transitions between the
eigenstates of a spin 1/2 system.

The spectral line shape is determined via
the corresponding stochastically averaged propagator of the Kubo oscillator
as \cite{Kubo62}
\begin{equation}\label{shape}
I (\omega)=\frac{1}{\pi}\lim_{\eta \to +0} {\rm Re}
[ \tilde S(-i\omega+\eta)]\;.
\end{equation}
Note that the limit $\eta\to +0$ in Eq. (\ref{shape}) is necessary
for the regularization of the corresponding integral in the quasi-static
limit $\langle \tau_j\rangle \to \infty$.
Upon identifying $\mathcal{ L}[\xi_k]$ with $-\epsilon_{k}$ in Eq. (\ref{final})
we end up with
\begin{eqnarray}\label{propN}
\langle \tilde S(s) \rangle & = & \sum_{k}\frac{p_k^{st}}{s-i\epsilon_k}
-\sum_{k}\frac{1-\tilde \psi_k(s-i\epsilon_k)}{(s-i\epsilon_k)^2}
\frac{p_k^{st}}{\langle \tau_k\rangle }\\
\nonumber
& + &
\sum_{n,l,m}\frac{1-\tilde \psi_l(s-i\epsilon_l)}{s-i\epsilon_l}
\Big (\frac{1}{I-P\tilde D(s)}\Big )_{lm}p_{mn}
\frac{1-\tilde \psi_n(s-i\epsilon_n)}
{s-i\epsilon_n}\frac{p_n^{st}}{\langle \tau_n\rangle }
\end{eqnarray}
with $\tilde D_{nm}(s)=\delta_{nm}\tilde \psi_m(s-i\epsilon_m)$
\footnote{Note that the formal solution of another important problem, namely the
the (first order) relaxation kinetics with a
fluctuating rate, $\dot p(t)=-\Gamma[\xi(t)] p(t)$
follows immediately from (\ref{propN}) upon substitution
$\epsilon_j\to i\Gamma_j$, see in Ref.  \cite{JCP05} 
containing some nontrivial
non-Markovian examples.}.
The corresponding
line shape follows  from Eq. (\ref{propN}) by virtue of
Eq. (\ref{shape}). This result presents a non-Markovian
generalization of the pioneering result by Kubo \cite{Kubo62} for
arbitrary $N$-state
discrete Markovian processes. This generalization
applies to
arbitrary non-exponential RTDs $\psi_k(\tau)$, or, equivalently, in
accordance with Eq. (\ref{reduce}) also for time-dependent transition
rates $k_{ij}(\tau)$. This generalization was put forward originally  in Ref.
\cite{Goychuk04} for a
particular case, $p_j^{st}=\langle \tau_j \rangle/
\sum_k\langle \tau_k \rangle$, which corresponds to an ergodic
process with uniform mixing (meaning that in a long time limit each state $j$
is visited equally often).

Next we
apply this result to the case of two-state
non-Markovian noise with $p_{12}=p_{21}=1$ and
$p_{1,2}^{st}=\langle \tau_{1,2}\rangle/[\langle \tau_{1}\rangle
+\langle \tau_{2}\rangle]$.  Eq. (\ref{propN}) then
yields:
\begin{eqnarray}\label{prop2}
\langle \tilde S(s)\rangle & = & \sum_{k=1,2}\frac{1}{s-i\epsilon_k}
\frac{\langle \tau_k \rangle}{\langle \tau_1 \rangle+
\langle \tau_2 \rangle}
 + \frac{(\epsilon_1-\epsilon_2)^2}{(\langle \tau_1 \rangle+
\langle \tau_2 \rangle)(s-i\epsilon_1)^2(s-i\epsilon_2)^2}\nonumber \\
&\times &
\frac{[1-\tilde \psi_1(s-i\epsilon_1)][1-\tilde \psi_2(s-i\epsilon_2)]}
{1-\tilde \psi_1(s-i\epsilon_1)\tilde \psi_2(s-i\epsilon_2)}\; .
\end{eqnarray}
 With (\ref{prop2}) substituted into (\ref{shape}) one finds the result
 for the corresponding
spectral line shape which matches that  presented
recently in Ref. \cite{JungBarkSilb}
derived therein by use of a  different method.  Moreover,
in the simplest case of Markovian two-state fluctuations
with $\tilde \psi_{1,2}(s)=1/(1+\langle \tau_{1,2}\rangle s)$
and with zero mean, $\langle \xi(t)\rangle =\langle \tau_1\rangle
\epsilon_1 +\langle \tau_2\rangle
\epsilon_2=0$, this
result simplifies further to  read
\begin{eqnarray}\label{Markov}
\langle \tilde S(s)\rangle =\frac{s+2\chi}{s^2+2\chi s +\sigma^2}\;.
\end{eqnarray}
In (\ref{Markov}),
$\sigma=\sqrt{\langle \xi^2(t)\rangle}=|\epsilon_2-\epsilon_1|
\sqrt{\langle \tau_1\rangle \langle \tau_2\rangle}/
(\langle \tau_1\rangle + \langle \tau_2\rangle)$ denotes the root mean
squared (rms) amplitude of fluctuations. Moreover,
$\chi=\nu/2+i\sigma\sinh(b/2)$
denotes a complex frequency parameter, where $\nu=1/\langle \tau_1\rangle+
1/\langle \tau_2\rangle$ is the inverse of the autocorrelation time
of the considered process. Its
 autocorrelation function 
 reads $\langle \xi(t)\xi(t')\rangle=\sigma^2\exp(-\nu|t-t'|)$.
Furthermore, $b=\ln(\langle \tau_1\rangle/
\langle \tau_2\rangle)=\ln|\epsilon_2/\epsilon_1|$
is an asymmetry parameter.
The spectral line shape corresponding to (\ref{Markov}) has been
 obtained by Kubo, reading
 \cite{Kubo62,PRE94},
\begin{eqnarray}\label{kuboline}
I(\omega)=\frac{1}{\pi}\frac{\sigma^2\nu}
{(\omega+\epsilon_1)^2(\omega+\epsilon_2)^2+\omega^2\nu^2}.
\end{eqnarray}
Moreover, the expression (\ref{Markov}) can  readily be inverted into
its time domain.
Note that the corresponding averaged propagator
$\langle S(t)\rangle $ of
Kubo oscillator \cite{PRL98}, i.e.,
\begin{eqnarray}\label{propKubo}
\langle S(t)\rangle=e^{-\chi t}\Big [\cos(\sqrt{\sigma^2-\chi^2}t)+
\frac{\chi}{\sqrt{\sigma^2-\chi^2}}\sin(\sqrt{\sigma^2-\chi^2}t)\Big],
\end{eqnarray}
is complex-valued  when the process $\xi(t)$ is asymmetric, i.e. $b\neq 0$. This
is in accordance with the asymmetry of the corresponding spectral line shape,
$I(-\omega)\neq I(\omega)$.
Derived in a different context \cite{ReilSkin}  for the case of  a two-state Markovian
process with
a nonvanishing mean and in quite different notation
an expression equivalent to (\ref{propKubo}) is
known in the theory of single-molecule spectroscopy
\cite{ReilSkin,GevaSkinner,Barkai}.
For a  symmetric dichotomous
process (with $b=0$) Eq. (\ref{propKubo}) reduces to the expression
(6.10) (with $\omega_0=0$) in Ref.
\cite{KampenBook}.

\section{Averaged dynamics of a two-level quantum systems
driven by  two-state
noise  \label{TLSclassic}}

Our non-Markovian stochastic theory of quantum relaxation
can be further
exemplified for the  relevant case of a two-state quantum system, reading
\begin{eqnarray}
\label{eq:TLS0}
H(t)=E_1|1\rangle\langle 1|+
E_2|2\rangle\langle 2|+\frac{1}{2}\hbar\xi(t)
(|1\rangle \langle 2|+|2\rangle \langle 1|),
\end{eqnarray}
which is driven by a two-state non-Markovian stochastic
noise $\xi(t)=\{\Delta_+,\Delta_-\}$ with corresponding RTDs $\psi_+(\tau),\psi_-(\tau)$
and the stationary state probabilities $p_{\pm}^{st}=\langle \tau_{\pm}\rangle/
[\langle \tau_{+}\rangle + \langle \tau_{-}\rangle]$.
This noise assumes  the normalized
stationary autocorrelation function \\
$k(t):=
\langle \delta\xi(t)\delta\xi(0)\rangle_{st}/
\langle [\delta\xi]^2\rangle_{st}$
where $\delta \xi(t):=\xi(t)-\langle \xi\rangle_{st}$ is temporal fluctuation.
Its Laplace-transform  reads \cite{Strat,LowenTeich,PRL03,PRE04}
\begin{eqnarray}\label{laplace-corr}
\tilde k(s)=
\frac{1}{s}-\left(\frac{1}{\langle\tau_+\rangle}+
\frac{1}{\langle\tau_-\rangle}\right)\frac{1}{s^2}
\frac{(1-\tilde\psi_+(s))(1-\tilde\psi_-(s))}{1-\tilde\psi_+(s)
\tilde\psi_-(s)}\;.
\end{eqnarray}
This dichotomic noise possesses the   power spectrum $S_N(\omega)$:
\begin{equation}\label{spower}
S_N(\omega)=\frac{2(\Delta_+-\Delta_-)^2}{\langle\tau_+\rangle+
\langle\tau_-\rangle}
\frac{1}{\omega^2}{\rm Re}\left[\frac{(1-\tilde\psi_+(i\omega))
(1-\tilde\psi_-(i\omega))}{1-\tilde\psi_+(i\omega)
\tilde\psi_-(i\omega)} \right].
\end{equation}

It causes (dipole) transitions between two states,
$|1\rangle$ and $|2\rangle$, and is zero on average (a first interpretation).
A different interpretation of the considered dynamics can also be given
when $\xi(t)$ does not vanish on average. Then,
for $\Delta_+>\Delta_->0$, we are dealing with a quantum tunneling
dynamics with a fluctuating tunneling matrix element, e.g. due to a
fluctuating tunneling barrier.

For the considered case of a two state non-Markovian process with
$p_{11}=p_{22}=0$, $p_{12}=p_{21}=1$ the general
result in Eq. (\ref{final}) can be simplified further. After
some cumbersome operator algebra we obtain:
\begin{eqnarray}\label{newfinal}
\langle \tilde S(s)\rangle=p_{+}^{st}\tilde S_{+}(s)+p_{-}^{st}\tilde S_{-}(s)
-\frac{1}{\langle \tau_{+}\rangle +\langle \tau_{-}\rangle}
\{ \tilde C_+(s)+\tilde C_-(s) \nonumber \\-[\tilde A_{+}(s)
\tilde B_{-}(s)+\tilde A_{-}(s)][ I-\tilde B_{+}(s)
\tilde B_{-}(s)]^{-1} \tilde A_{+}(s) \\
- [\tilde A_{-}(s)
\tilde B_{+}(s)+\tilde A_{+}(s)][I-\tilde B_{-}(s)
\tilde B_{+}(s)]^{-1} \tilde A_{-}(s)\} \nonumber,
\end{eqnarray}
where $\tilde S_{\pm}(s)$ denotes the Laplace-transform of the
propagator $S_{\pm}(t)=\exp(-i\mathcal{L}_{\pm}t)$ with
$\mathcal{L}_{\pm}:=\mathcal{L}[\Delta_{\pm}]$ corresponding to the {\it fixed} noise value $\xi=\Delta_+$
and $\xi=\Delta_-$, correspondingly. Furthermore, $\tilde C_{\pm}(s)$
is given by Eq. (\ref{aux1new}), $\tilde C_{\pm}(s)\equiv
\tilde C_{\pm\pm}(s)$ and
\begin{equation}\label{aux1ad}
\tilde A_{\pm}(s):=
\int_{0}^{\infty}\Phi_{\pm}(\tau)\exp[-(s+i \mathcal{ L}_{\pm})\tau]
d\tau,
\end{equation}
\begin{equation}\label{aux2ad}
\tilde B_{\pm}(s):=
\int_{0}^{\infty}\psi_{\pm}(\tau)\exp[-(s+i \mathcal{ L}_{\pm})\tau]
d\tau\;.
\end{equation}
A quantum evolution characterized in Eqs. (\ref{newfinal})-(\ref{aux2ad})
by the Liouville operators
$\mathcal{ L}_{\pm}$ does  belong to a rather broad class
(e.g.,  for quantum systems with a finite number of states)
and is not merely restricted
to the case of two-state quantum dynamics in Eq. (\ref{eq:TLS0}).

The archetype  model in Eq. (\ref{eq:TLS0}) does exhibit a rich behavior.
In particular, it opens a doorway to
study the problem of quantum decoherence of a two-state atom
under the influence of
two-state ``$1/\omega^{\alpha}$''
noise that  exhibits long range time-correlations
with a power law decay (for $\psi(\tau)$ possessing a long-time
algebraic tail, $\psi(\tau)\propto 1/\tau^{3-\alpha}$,
$0<\alpha<1$)\cite{LowenTeich,PRL03,PRE04}.
Therefore, this model constitutes a prominent problem of 
general interest. Moreover, it relates
to recent activities that involve decoherence studies for  
solid state quantum computing \cite{Wilhelm}.
It is convenient
to express the Hamiltonian (\ref{eq:TLS0}) in terms of Pauli
matrices, $\hat \sigma_z:=|1\rangle \langle 1|-|2\rangle \langle 2|$,
$\hat \sigma_x:=|1\rangle \langle 2|+|2\rangle \langle 1|$,
$\hat \sigma_y:=i(|2\rangle \langle 1|-|1\rangle \langle 2|)$ and
the unit matrix $\hat I$,
\begin{eqnarray}
\label{eq:TLS1}
H(t)=\frac{1}{2}\hbar\epsilon_0 \hat \sigma_z+\frac{1}{2}\hbar\xi(t)
\hat \sigma_x + \frac{1}{2}(E_1+E_2)\hat I,
\end{eqnarray}
where $\epsilon_0=(E_1-E_2)/\hbar$.
The dynamics of the density matrix of the quantum
two-state quantum system is then obtained
as $\rho(t)=\frac{1}{2}[\hat I +\sum_{i=x,y,z}\sigma_i(t)\hat \sigma_i]$
with components  $\sigma_i(t)=
{\rm Tr}(\rho(t)\hat\sigma_i)$.
This latter dynamics evolves on a Bloch sphere of unit radius (i.e.,
the corresponding (scaled) magnetic moment is conserved, i.e.,
\footnote{This means that each and every
stochastic trajectory runs on the Bloch
sphere. The stochastically {\it averaged} Bloch vector dynamics $\langle\vec \sigma (t)\rangle$
is, however, ''dissipative``; i.e. $|\langle\vec \sigma (t)\rangle|\leq 1$,
because $\langle \sigma_i(t)\rangle^2\leq \langle \sigma_i^2(t)\rangle$.
Thus, the averaged density matrix $\langle \rho(t)\rangle$
remains always positive  in the considered model, independently
of the particular model used for the stochastic force $\xi(t)$.},
$|\vec \sigma (t)|=1$). Its rate of change obeys,
\begin{eqnarray}
\label{eq:Bloch}
 \dot \sigma_x(t) &=& -\epsilon_0 \sigma_y(t), \nonumber \\
 \dot \sigma_y(t) &=& \epsilon_0 \sigma_x(t)-\xi(t) \sigma_z(t), \\
 \dot \sigma_z(t) &=& \xi(t) \sigma_y(t) \;,\nonumber
\end{eqnarray}
or $\dot{\vec{\sigma}}(t)=\hat F[\xi(t)]\vec \sigma(t)$ in the vector
form, where
\begin{eqnarray}
\hat F[\xi(t)]=
\left( \begin{array}{ccc}
0 & -\epsilon_0 & 0 \\
\epsilon_0 & 0 & -\xi(t)  \\
0 & \xi(t) & 0 \end{array} \right)\;\;.
\end{eqnarray}
The above  theory  can  readily be applied to the noise
 averaging of a 3-dimensional system of linear differential equations
(\ref{eq:Bloch}) over arbitrary {\it stationary} realizations of $\xi(t)$
with the obvious formal substitution $-i \mathcal{ L}[\xi(t)]\to F[\xi(t)]$.
Towards this goal, we  represent the  propagators
$\hat S_{\pm}(t)=\exp(\hat F[\Delta_{\pm}]t)$
for the fixed static values of noise $\xi=\Delta_+$ and $\xi=\Delta_-$
as matrix expansions over the eigen-modes of evolution
$\exp(i\lambda_{\pm}^{(k)}t)$, with  $\lambda_{\pm}^{(0)}=0$, $\lambda^{(1)}_{\pm}=
\Omega_{\pm}$,  $\lambda^{(2)}_{\pm}=-
\Omega_{\pm}$, where $\Omega_{\pm}=\sqrt{\epsilon_0^2+\Delta_{\pm}^2}$
are the eigen-frequencies of coherent quantum oscillations for
constant $\xi=\Delta_{\pm}$. It reads,
\begin{eqnarray}
\hat S_{\pm}(t)=\sum_{k=0,1,2}\hat R_{\pm}^{(k)}\exp(i\lambda_{\pm}^{(k)}t),
\end{eqnarray}
where
\begin{eqnarray}\label{Rexplicit}
\hat R_{\pm}^{(0)}& = &\frac{1}{\Omega_{\pm}^2}
\left( \begin{array}{ccc}
\Delta_{\pm}^2 & 0 & \epsilon_0\Delta_{\pm}  \\
0 & 0 & 0  \\
\epsilon_0\Delta_{\pm} & 0 & \epsilon_0^2 \end{array} \right),\nonumber \\
\hat R_{\pm}^{(1)}& = &[\hat R_{\pm}^{(2)}]^*=\frac{1}{2}
\left( \begin{array}{ccc}
\frac{\epsilon_0^2}{\Omega_{\pm}^2} & i
\frac{\epsilon_0}{\Omega_{\pm}} &
-\frac{\epsilon_0\Delta_{\pm}}{\Omega^2_{\pm}}  \\
  -i
\frac{\epsilon_0}{\Omega_{\pm}}& 1 &  i
\frac{\Delta_{\pm}}{\Omega_{\pm}}  \\
 -\frac{\epsilon_0\Delta_{\pm}}{\Omega^2_{\pm}}&  -i
\frac{\Delta_{\pm}}{\Omega_{\pm}}  &\frac{\Delta_{\pm}^2}{\Omega^2_{\pm}}
\end{array} \right), \;\sum_{k=0,1,2}\hat R_{\pm}^{(k)}=\hat I.
\end{eqnarray}
The corresponding Laplace-transformed matrices entering Eq. (\ref{newfinal})
are represented as
\begin{eqnarray}\label{complete}
\tilde S_{\pm}(s)& = &\sum_{k=0,1,2}\frac{\hat R_{\pm}^{(k)}}{
s-i\lambda_{\pm}^{(k)}}, \nonumber \\
\tilde A_{\pm}(s) & = &\sum_{k=0,1,2}\hat R_{\pm}^{(k)}\frac{1-
\tilde \psi_{\pm}(s-i\lambda_{\pm}^{(k)})}{
s-i\lambda_{\pm}^{(k)}}, \nonumber \\
\tilde B_{\pm}(s)& = &\sum_{k=0,1,2}\hat R_{\pm}^{(k)}
\tilde \psi_{\pm}(s-i\lambda_{\pm}^{(k)}), \\
\tilde C_{\pm}(s)& = &\sum_{k=0,1,2}\hat R_{\pm}^{(k)}\frac{1-
\tilde \psi_{\pm}(s-i\lambda_{\pm}^{(k)})}{
(s-i\lambda_{\pm}^{(k)})^2} \;. \nonumber
\end{eqnarray}
The results in Eqs. (\ref{newfinal}),(\ref{Rexplicit}), (\ref{complete})
form our  basis for further studies.

Let us evaluate it explicitly for the case of symmetric process,
$\psi_+(\tau)=\psi_-(\tau)=\psi(\tau)$, $\langle\tau_+\rangle=
\langle\tau_-\rangle=\langle\tau\rangle$, with zero mean value,
$\Delta_+=-\Delta_{-}=\Delta$. We consider the situation
where  the state ``1'' is populated initially with
the probability one: $\sigma_z(0)=1,\sigma_{x,y}(0)=0$.
Then, the Laplace-transform of the averaged difference of two populations, i.e.,
$\langle \sigma_z(t)\rangle=\langle \rho_{11}(t)\rangle-
\langle \rho_{22}(t)\rangle$ is denoted as
$\langle \tilde \sigma_z(s)\rangle=\langle \tilde S_{zz}(s)\rangle$.
After some cumbersome manipulations
using a computer
algebra system (MAPLE) to perform multiple matrix operations,
we end up with the following compact result:
\begin{eqnarray}
\label{eq:solution}
\langle \tilde \sigma_z(s)\rangle=\frac{s^2+\epsilon_0^2}{s(s^2+\Omega^2)}-
\frac{2\Delta^2}{\langle\tau\rangle s^2(s^2+\Omega^2)^2}\frac{\tilde A_{zz}(s)}
{\tilde B_{zz}(s)},
\end{eqnarray}
where
\begin{eqnarray}
  \label{eq:Azz}
  \tilde A_{zz}(s) & = &\epsilon_0^2 [1-\tilde\psi(s)]
\{ (\Omega^2-s^2)(1-\tilde\psi(s+i\Omega)
\tilde\psi(s-i\Omega))  \nonumber \\
& - &
2 i \Omega\, s\, [\tilde \psi(s+i\Omega)-\tilde\psi(s-i\Omega)]\}
\nonumber \\ & - & \Delta^2 s^2[1+\tilde\psi(s)]
[1-\tilde\psi(s+i\Omega)][1-\tilde\psi(s-i\Omega)],  \\
\tilde B_{zz}(s) &= &\epsilon_0^2[1-\tilde\psi(s)]
[1+\tilde\psi(s+i\Omega)]
[1+\tilde\psi(s-i\Omega)] \nonumber \\ & + &
\Delta^2  [1+\tilde\psi(s)] (1-\tilde\psi(s+i\Omega)
\tilde\psi(s-i\Omega)). \nonumber
\end{eqnarray}
Note that for the considered initial condition, we find
$\langle \sigma_x(t)\rangle=
\langle  \sigma_y(t)\rangle=0$ for all times because
$\langle \tilde S_{xz}(s)\rangle=\langle \tilde S_{yz}(s)\rangle=0$.
For $\epsilon_0=0$ the result in (\ref{eq:solution})-(\ref{eq:Azz})
reduces to one for
Kubo oscillator (\ref{prop2}) with identical $\psi_{1,2}(\tau)$. Moreover,
for the Markovian case, $\tilde\psi(s)=1/(1+\tau s)$, Eq.
(\ref{eq:solution}) reduces to
\begin{eqnarray}
\label{sigmaZ}
\langle \tilde \sigma_z(s)\rangle=\frac{s^2+2\nu s+ \nu^2+\epsilon_0^2}
{s^3+2\nu
  s^2+ (\Delta^2 + \epsilon_0^2 +\nu^2)s +\Delta^2 \nu},
\end{eqnarray}
where $\nu=2/\langle\tau\rangle$ is the inverse autocorrelation time.
 This latter result reproduces
the result for the averaged populations $\langle \tilde
\rho_{11}(s)\rangle=(1/s+\langle \tilde \sigma_z(s)\rangle)/2$
and $\langle \tilde
\rho_{22}(s)\rangle=(1/s-\langle \tilde \sigma_z(s)\rangle)/2$
in \cite{AverPouq,ShapLog1}.
The same result (\ref{sigmaZ}) can also be deduced from the
known solution
for the Markovian case driven by   asymmetric two-state noise
 \cite{PLA96} when specified to symmetric noise limit.

This population difference  possesses several remarkable
features: The asymptotic difference between populations is zero,
$\langle \sigma_z(\infty)\rangle=\lim_{s\to 0}(s
\langle \tilde \sigma_z(s)\rangle)=0$. In other words, the steady
state populations of both energy levels  equal  $1/2$, independent
of the energy difference $\hbar\epsilon_0$.
This result can be elucidated best in terms of a ``temperature''
$T_{\sigma}$ of the (pseudo-)spin system. This spin-temperature is formally introduced
by using for the  asymptotic distribution an Ansatz of the Boltzmann-Gibbs
form,
$\langle\rho_{nn}(\infty)\rangle = \exp[-E_n/k_BT_{\sigma}]/
\sum_{n}\exp[-E_n/k_BT_{\sigma}]$.
Then\footnote{This is the standard definition
of the temperature of a spin subsystem in nuclear magnetic
resonance and similar research areas  \cite{Slichter}. It
is used also to introduce the parlance of formally
{\it negative} temperatures.},
\begin{eqnarray}\label{Tsigma}
T_{\sigma}:=\frac{\hbar\epsilon_0}
{k_B\ln\left(\frac{\langle\rho_{22}(\infty)\rangle}
{\langle\rho_{11}(\infty)\rangle} \right)}\;
\end{eqnarray}
for two-level systems.
In accordance with this definition, the result of equal asymptotic populations,
$\langle\rho_{22}(\infty)\rangle=\langle\rho_{11}(\infty)\rangle=1/2$
can be interpreted in terms of an infinite temperature $T_{\sigma}=\infty$.
This constitutes a general finding: a purely stochastic bath corresponds to
an apparently infinite
temperature \cite{ReinekerBook,LindWest}. Thus, this stochastic
approach modeling  the relaxation
process in open quantum systems is suitable only
for sufficiently
high temperatures $k_BT\gg \hbar|\epsilon_0|$ \cite{ReinekerBook,LindWest}.
An asymmetry of unbiased stochastic fluctuations
does not impact this  conclusion, see in Ref. \cite{PLA96}.
Moreover, the relaxation to the steady state can be
either coherent, or incoherent,
depending  on the noise strength and the value of autocorrelation time. In particular,
an approximate analytical expression for the rate $k$ of incoherent
relaxation, $\langle\rho_{11}(t)\rangle=[1+\exp(-k t)]/2$,
has been obtained in a limit of small Kubo numbers, $K:=\Delta/\nu\ll 1$,
which corresponds to a weakly colored noise
\cite{KampenBook,FrischBrissaud2}.
This analytic result reads \cite{AverPouq,ShapLog1,PLA96}
\begin{equation}\label{resonance}
k=\frac{\Delta^2\nu}{\nu^2+\epsilon_0^2} \;.
\end{equation}
The rate exhibits a resonance feature versus $\nu$ at $\nu=\epsilon_0$.
A similar resonance feature
occurs also in the theory of nuclear magnetic resonance for  weakly colored
Gaussian noise \cite{Slichter}. Note that in Ref. \cite{PLA96}
this notable
result has been obtained for asymmetric fluctuations of the tunneling
coupling possessing a non-vanishing mean
value $\langle \xi(t)\rangle \neq 0$.
This corresponds to  a quantum particle transfer between
two sites of localization which are separated by
a fluctuating tunneling barrier. A related  problem with the inclusion of
quantum dissipation has been
elaborated in \cite{JCP95} within a stochastically
driven spin-boson model.

Yet another interesting solution can be  obtained for $\langle \tilde
\sigma_x(s)\rangle$ with the initial condition  reading $\sigma_x(0)=1$.
The Laplace transform of the solution reads
\begin{eqnarray}
\label{eq:solution2}
\langle \tilde \sigma_x(s)\rangle=\frac{s^2+\Delta^2}{s(s^2+\Omega^2)}-
\frac{2\Delta^2\epsilon_0^2\Omega^2}{\langle\tau\rangle s^2(s^2+\Omega^2)^2}
\frac{\tilde A_{xx}(s)}
{\tilde B_{xx}(s)},
\end{eqnarray}
where
\begin{eqnarray}
  \label{eq:Axx}
  \tilde A_{xx}(s) & = &[1-\tilde\psi(s)]
[1-\tilde\psi(s+i\Omega)][1-\tilde\psi(s-i\Omega)],  \\
\tilde B_{xx}(s) &= &\epsilon_0^2[1+\tilde\psi(s)][1-
\tilde\psi(s+i\Omega)]
[1-\tilde\psi(s-i\Omega)] \nonumber \\ & + &
\Delta^2  [1-\tilde\psi(s)] (1-\tilde\psi(s+i\Omega)
\tilde\psi(s-i\Omega)). \nonumber
\end{eqnarray}
The physical relevance of this solution (\ref{eq:solution2}) is as follows:
In the rotated pseudo-spin basis,
$\hat \sigma_x\to \hat \sigma_z$,
$\hat \sigma_z\to \hat \sigma_x$, $\hat \sigma_y\to \hat
\sigma_y$, this problem becomes mathematically
equivalent to the problem
of the delocalization of a quantum particle in a symmetric dimer with
a tunneling coupling $\epsilon_0$ under the influence of a
dichotomously fluctuating energy bias $\xi(t)$.
Therefore, it
describes the corresponding delocalization dynamics and, in
particular, allows one to determine whether this dynamics is
coherent or incoherent,
depending on the noise characteristics.

For the Markovian case, Eq. (\ref{eq:solution2}) reduces to
\footnote{The corresponding
dynamics also exhibits a resonance feature versus the inverse autocorrelation time $\nu$ within  certain
limits \cite{AnkerPech}.}
\begin{eqnarray}
\label{sigmaX}
\langle \tilde \sigma_x(s)\rangle=\frac{s^2+\nu s +\Delta^2}
{s^3+\nu
  s^2+ (\Delta^2 + \epsilon_0^2)s +\epsilon_0^2 \nu}.
\end{eqnarray}
Note that the denominators in Eq. (\ref{sigmaZ}) and Eq. (\ref{sigmaX})
are different\footnote{A notable feature is, however, that
both corresponding secular cubic equations have the
{\it same} discriminant,
$D(\Delta,\nu,\epsilon_0)=0$,
separating the domains of complex and real roots. Hence,
the transition from a coherent relaxation
(i.e. complex roots are present) to an incoherent relaxation
(i.e. only real roots are obtained) occurs
at the same values of noise parameters,
independently of the initial conditions.
The corresponding phase diagram separating regimes of coherent
and incoherent relaxation (judging from the above criterion)
has been devised in \cite{AnkerPech}.
It must be  kept in mind, however, that the weights of the corresponding
exponentials are also of importance for the characteristic relaxation dynamics.
These weights do depend  on the initial conditions.}.
In the more general case of asymmetric Markovian noise,
the corresponding denominator is
 a polynomial of 6-th order in $s$, see in \cite{PLA96}.
In the considered case of symmetric noise it factorizes
into the product of two  polynomials of 3-rd order,
namely into those in the denominators of Eq. (\ref{sigmaZ}) and Eq. (\ref{sigmaX}).
Thus, for a general initial condition the relaxation of a two-level
quantum system exposed to a two-state Markovian field
involves six exponential terms. As a matter of fact,
this seemingly simple, {\it exactly} solvable model can exhibit
an unexpectedly complex behavior even in the  Markovian case
of a colored two-state noise.
However, for certain initial
conditions, as exemplified above, the general solution being a fraction of two polynomials
of $s$  simplifies to the results in Eq. (\ref{sigmaZ}) and Eq. (\ref{sigmaX}).

In a general case of non-Markovian noise, the analytical solutions
in Eqs. (\ref{eq:solution}) and (\ref{eq:solution2}) can be inverted numerically
to the time domain by use  of a  numerical Laplace inversion procedure
such as the one detailed in Ref. \cite{Stehfest} using a computer algebra
implementation with arbitrary digital precision \cite{Valko}. A quasi-analytical
inversion is, however, still possible in specific non-Markovian cases.

\section{A  simple non-Markovian case}

We consider the following basic case of non-Markovian
noise described by a bi-exponential
RTD
\begin{eqnarray}
\psi(\tau)=\theta \alpha_1\exp(-\alpha_1 \tau)+(1-\theta)
\alpha_2\exp(-\alpha_2 \tau),
\end{eqnarray}
where $\alpha_{1,2}$ denote two transition rates which can be realized
with probabilities $\theta$ and $1-\theta$, correspondingly. The considered
noise possesses the mean residence time
\begin{eqnarray}
\langle \tau\rangle =\theta/\alpha_1+(1-\theta)/\alpha_2
\end{eqnarray}
and the mean autocorrelation time
$\tau_{corr}:=\int_{0}^{\infty} |k(t)|dt$ reading with  $k(t)\geq 0$,
\begin{eqnarray}
\tau_{corr}=C_V^2 \tau_{corr}^{(M)},
\end{eqnarray}
where $C_V=\sqrt{\langle \tau^2\rangle -\langle \tau\rangle^2}/
\langle \tau\rangle$ is the coefficient of variation of the RTD
\cite{PRL03,PRE04} and $\tau_{corr}^{(M)}=\langle \tau \rangle/2$
is the autocorrelation time of Markovian process possessing the
same mean residence time $\langle \tau \rangle$.
The ratio $\tau_{corr}/\tau_{corr}^{(M)} =C_V^2$
 serves as a  convenient
quantifier for non-Markovian effects. For this basic
non-Markovian case considered, we find that  $C_V^2$ can be large, see in Ref.
\cite{JCP05} for details. For example, in the limit
$\zeta=\alpha_1/\alpha_2\ll \theta\ll 1$, $C_V^2\approx 2/\theta \gg 1$.
Such a noise has distinct bursting features \cite{JCP05}.

For this noise, Eq. (\ref{eq:solution}) reduces to a rational
function of $s$,
\begin{eqnarray}\label{ND1}
\langle \tilde \sigma_z(s)\rangle=\frac{N_{zz}(s)}{D_{zz}(s)},
\end{eqnarray}
where $N_{zz}(s)$ and $D_{zz}(s)$ are specific polynomials of 5-th and 6-th orders,
correspondingly.
The explicit form of these polynomials is not given here since it
does not provide much physical insight.

The inversion of (\ref{ND1}) to the time domain reads
\begin{eqnarray}\label{inv1}
\langle  \sigma_z(t)\rangle=\sum_{k=1}^{6}\frac{N_{zz}(r_k)}
{D_{zz}^{'}(r_k)}\exp(r_k t),
\end{eqnarray}
where $r_k$ are the complex roots (${\rm Re}\;r_k < 0$) of $D_{zz}(s)=0$
and $D_{zz}^{'}(s):=dD_{zz}(s)/ds$.  Even if $r_k$ cannot be determined
analytically (except for the case $\epsilon_0=0$, where the considered
rational function can be simplified to the ratio of two polynomials
of the third and fourth orders, correspondingly),
they can be  found numerically for any set of
parameters entering the problem. The problem is thereby quasi-analytically
solvable.
The same is valid for taking more terms in the exponential 
expansion of $\psi(\tau)$.
The whole scheme can be easily
implemented using a computer algebra system like MAPLE or MATHEMATICA.
In this respect it is pertinent to note that a power law dependence
can be well approximated by an expansion of the type $\psi(\tau)=\sum_{i}c_i\alpha_i
\exp(-\alpha_i \tau)$. Indeed, a power law extending
over $n$ decades can be satisfactory approximated  already by an $n$-term
expansion with properly
scaled $\{\alpha_i\}$ and $\{c_i\}$ \cite{Metzler}; see also a
practical example of a 6-exponential fitting
in \cite{Sansom}. Moreover, the numerical procedures
of the Laplace-transform inversion like one in \cite{Valko}
can be implemented. An analytical inversion of Eq. (\ref{eq:solution2})
to the time-domain has the form similar to Eq. (\ref{inv1}),
but with different
polynomials $N_{xx}(s)$ and $D_{xx}(s)$ of 5-th and 6-th order, correspondingly.

\subsection{Numerical Results}

\begin{figure}
\begin{center}
\includegraphics[width=12cm]{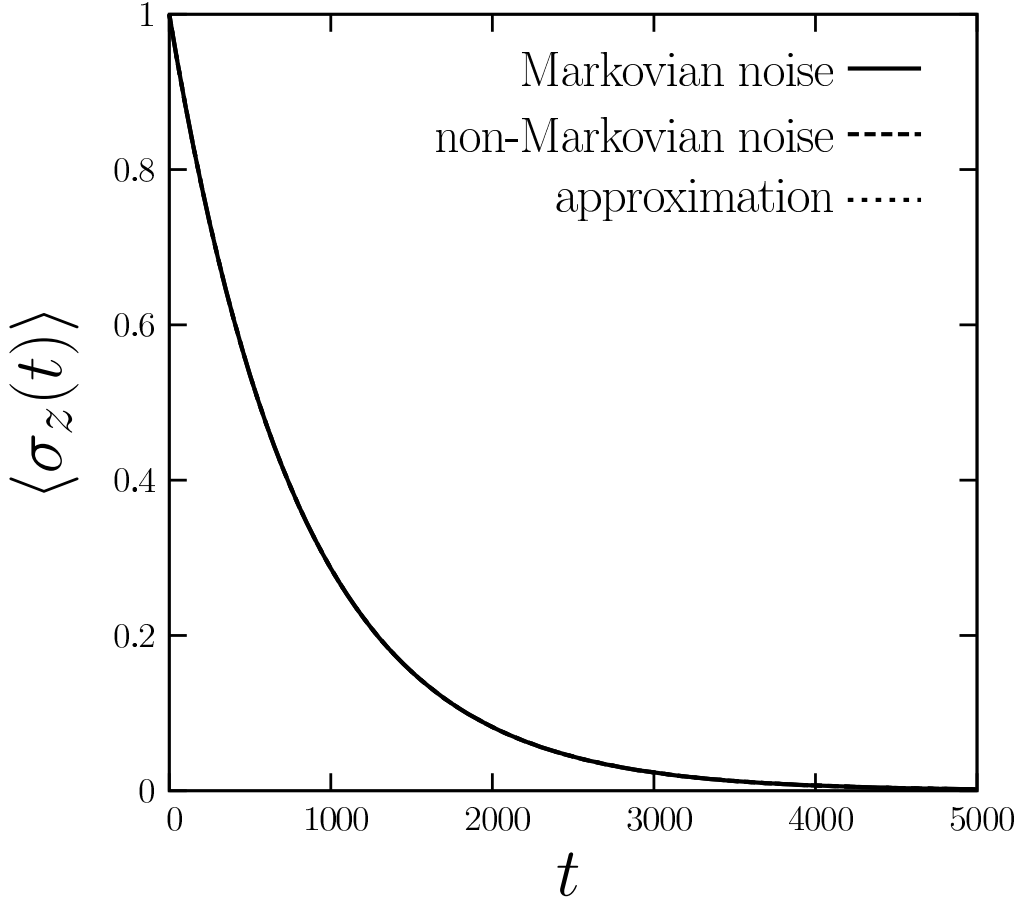}
\end{center}
\vspace{1cm}
\caption{Quantum relaxation in the perturbative regime. All three
curves practically coincide on the considered time
scale, see text for details.}
\label{Fig1}
\end{figure}

We consider first the situation with a small Kubo number
$K=\Delta\tau_{corr}\ll 1$ where an approximate analytical result,
$\langle \sigma_z(t)\rangle=\exp(-kt)$, is available with
the rate $k$ in Eq. (\ref{resonance}) \cite{AverPouq,PLA96}.
This result can also be obtained using the cumulant expansion method
\cite{KampenBook}. It does not depend on whether is the
stochastic process under consideration is Markovian or not. This
is illustrated with Fig. \ref{Fig1}. The following parameters are used
in calculations presented in Fig. \ref{Fig1}
(units are arbitrary): energy difference between levels
$\epsilon_0=1$, noise amplitude $\Delta=0.5$. Furthermore, for
the Markovian two-state noise we have chosen:
$\alpha_1=100$ (here $\theta=1$) and for the non-Markovian one:
$\alpha_1=100$, $\alpha_2=2000$,
$\theta\approx 0.05263157894$.
Both noises possess the same autocorrelation time $\tau_{corr}=0.005$
while the average residence times differ by an order of magnitude;
the non-Markovian parameter  is $C_V^2=10$.
On the characteristic time scale of quantum relaxation the relaxation
process in both cases is excellently described by the approximate
analytical result given above. All three lines practically coincide in
Fig. \ref{Fig1}. In this case, the perturbation theory in the small Kubo
number $K$ works very well.

It is interesting
to note that the initial decay of populations is
(beyond the discussed
exponential approximation) always Gaussian, i.e.
$\ln \langle \sigma_z(t)\rangle \propto -t^2$ at $t\to 0$.
This ``Gaussian'' regime
can be, however, very short and, therefore, it is not readily visible on
the characteristic time scale of the relaxation like in Fig.\ref{Fig1}.

\begin{figure}
\begin{center}
\includegraphics[width=12cm]{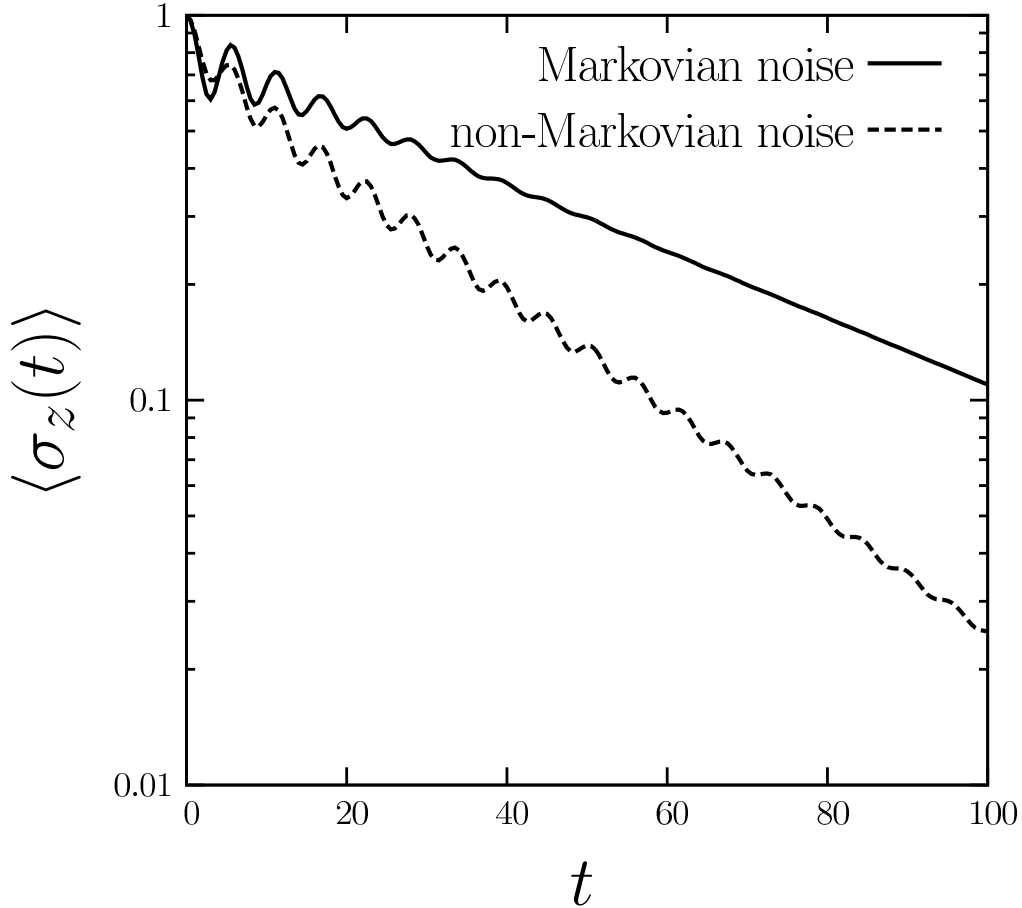}
\vspace{1cm}
\end{center}
\caption{Quantum relaxation in the non-perturbative regime of a large
Kubo number, $K=10$. The influence of Markovian and non-Markovian
noises with the same rms and mean autocorrelation time are distinctly
different. Coherent features are more pronounced
for the non-Markovian noise
case. Note that in the latter case the overall relaxation time is
shorter.}
\label{Fig2}
\end{figure}

In the opposite limit, $K\gg 1$, the distinction between the influence
of Markovian and non-Markovian noises possessing the  same (mean)
autocorrelation
time becomes rather distinct, cf. Fig. \ref{Fig2}. Here, the following
noise parameters are used: $\Delta=0.5$
$\alpha_1=0.05$, $\theta=1$ (for the Markovian noise); $\alpha_1=0.05$
$\alpha_2=1$,  $\theta\approx 0.05263157894$ (for the non-Markovian
noise). In both cases, the autocorrelation time is the same
$\tau_{corr}=10$, while the mean residence times are quite different:
$\langle\tau\rangle=20$ for the Markovian case and
$\langle\tau\rangle=2$ for the non-Markovian case.

\section{The manifest non-Markovian case}

For a situation in which  the autocorrelation time of noise $\tau_{corr}$ diverges,
the corresponding Kubo number $K=\Delta\tau_{corr}$ is infinite
and any perturbation theory is doomed to fail. The theory developed herein, however,
allows to obtain convergent numerical results by
performing the inverse Laplace transformation numerically. We illustrate
the strength of our theory with the following  example of manifest non-Markovian noise
with an  RTD given by
\begin{eqnarray}\label{PL}
\tilde \psi(s)=\frac{1}{1+s\langle\tau\rangle g_{\alpha}(s\tau_d)},
\end{eqnarray}
where
\begin{eqnarray}\label{PLaux}
g_{\alpha}(z)=\frac{\tanh(z^{\alpha/2})}{z^{\alpha/2}}
\end{eqnarray}
with $0<\alpha<1$. Here, $\langle\tau\rangle$ in Eq. (\ref{PL})
is the mean residence time and $\tau_d$ is a time constant presenting
an additional parameter of the distribution. For $\tau_d=0$, an exponential
distribution is restored.
The properties of the RTD determined by Eqs. (\ref{PL}), (\ref{PLaux})
are discussed in detail
in \cite{PREfrac}. This RTD  thus possesses  three parameters only, but it exhibits
an interesting repertoire of effects. In particular, it
can encompass up to three different interchanging power law
regimes with the asymptotic power law assuming the form
$\psi(\tau)\propto 1/\tau^{2+\alpha}$, $\tau\to\infty$ \cite{PREfrac}.

The corresponding noise has $1/\omega^{1-\alpha}$ feature in its power
spectrum at $\omega\to 0$. Its mean autocorrelation time $\tau_{corr}$ diverges,
$\tau_{corr}=\infty$. Therefore, the Kubo number $K$ is formally infinite and a corresponding
perturbation theory is questionable. Our theory delivers, however, {\it exact}
result for the Laplace-transformed noise-averaged quantum relaxation
of the excited level population. Moreover, the mean relaxation
time can be formally defined
as $\tau_{rel}:=\int_{0}^{\infty}\langle \sigma_z(t)\rangle d t$. With
Eq. (\ref{eq:solution}), this time follows as
$\tau_{rel}=\lim_{s\to 0}\langle \tilde \sigma_z(s)\rangle$ for
any  function $\tilde \psi(s)$.
Moreover, the asymptotic character of relaxation dynamics as $t\to \infty$ can
be found by using the Tauberian theorems of the Laplace-transform method
and small-$s$ expansion of $\tilde \psi(s)$ reading (in leading terms)
\begin{eqnarray}\label{leading}
\tilde \psi(s)\approx 1-\langle \tau\rangle s + A s^{1+\alpha}
\end{eqnarray}
for any $\psi(\tau)$ possessing the finite mean value $\langle \tau\rangle$
and the long-time algebraic tail $\psi(\tau)\propto 1/\tau^{2+\alpha}$.
For the particular case in Eq. (\ref{PL}), $A=\langle \tau\rangle
\tau_d^{\alpha}/3$.

Using (\ref{leading}) in Eq. (\ref{eq:solution}), after some algebra
we obtained in the leading order:
\begin{eqnarray}
\langle \tilde \sigma_z(s)\rangle\sim \frac{A\epsilon_0^2}{\Omega^2
\langle\tau\rangle} s^{\alpha-1},\;\; s\to 0.
\end{eqnarray}
Several remarkable results follow readily. First, $\langle
\sigma_z(\infty)\rangle=\lim_{s\to 0}[s\langle
\tilde \sigma_z(s)]=0$, i.e.
the general drawback of the stochastic Liouville approach is preserved. Second, the use
of a Tauberian theorem \cite{Feller} in the above equation  yields:
 \begin{eqnarray}\label{asymptotics0}
\langle \sigma_z(t)\rangle \sim \frac{A\epsilon_0^2}{\Omega^2\Gamma(1-\alpha)
\langle\tau\rangle} \frac{1}{t^{\alpha}},\;\;\; t\to \infty.
\end{eqnarray}
For the case in (\ref{PL}),  this latter equation is modified as
 \begin{eqnarray}\label{asymptotics}
\langle \sigma_z(t)\rangle \sim \frac{\epsilon_0^2}
{3\Omega^2\Gamma(1-\alpha)} \left (\frac{\tau_d}{t}\right)^{\alpha},\;\;\; t\to \infty.
\end{eqnarray}
Remarkably, this result does not depend on the mean residence time
$\langle \tau\rangle$. The tail of the relaxation curve clearly exhibits a power law, $\langle \sigma_z(t)\rangle
\propto 1/t^{\alpha}$.

\begin{figure}
\begin{center}
\includegraphics[width=10cm]{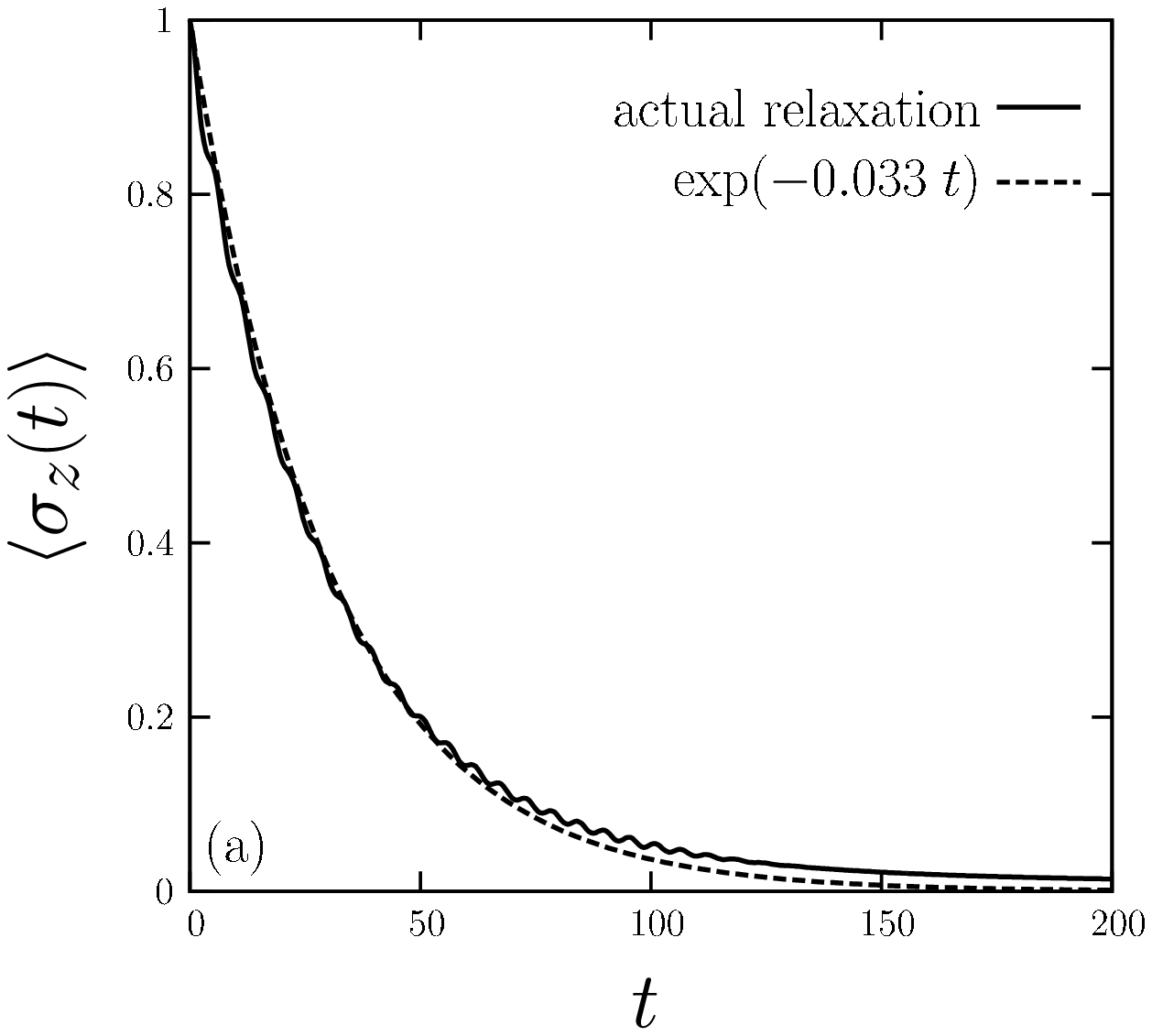}
\vspace{1cm}
\vfill
\includegraphics[width=10cm]{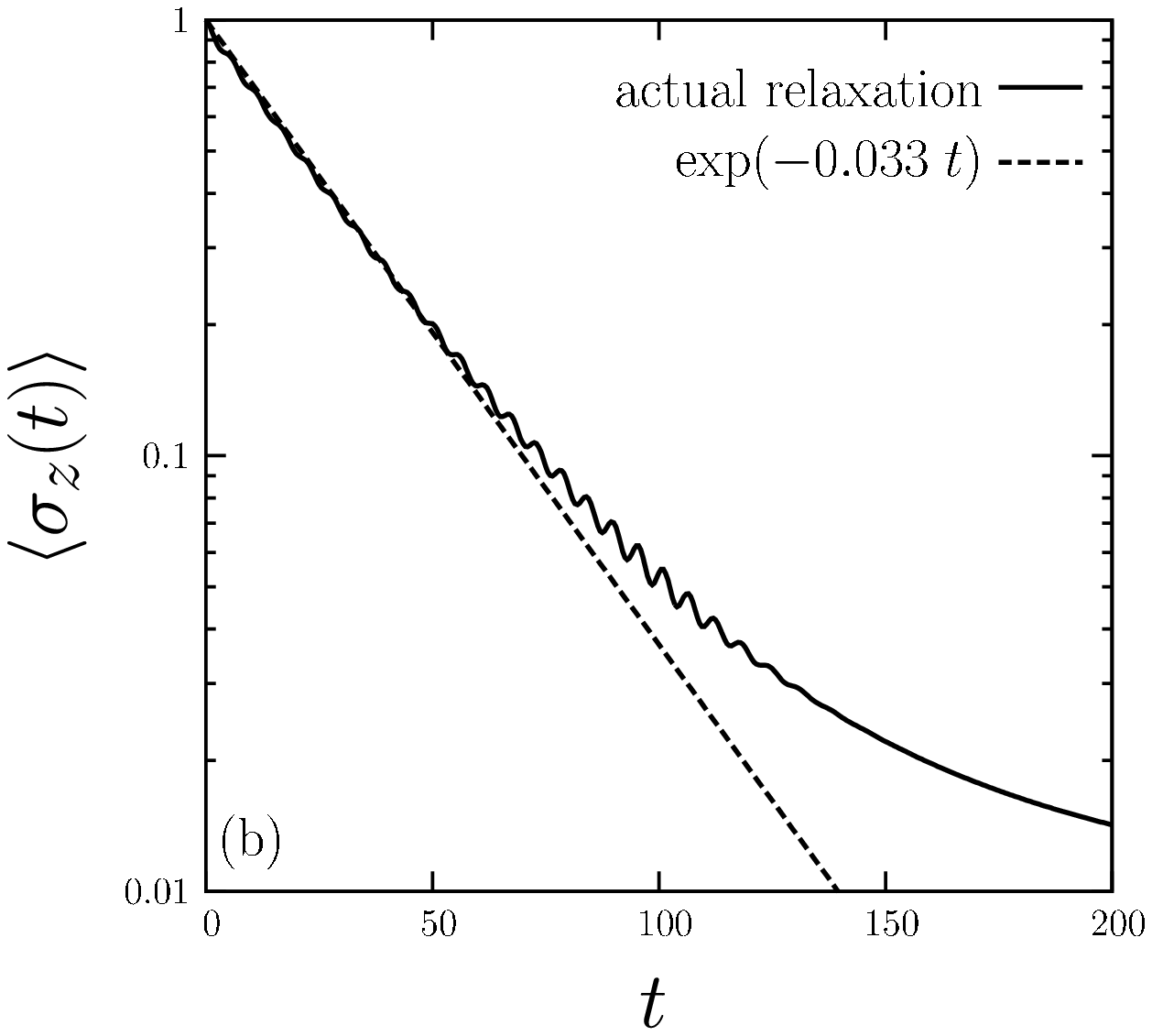}
\vspace{1cm}
\end{center}
\caption{(a) Quantum relaxation under the influence
of non-Markovian two state noise with extreme long time correlations.
The parameters of this manifest  non-Markovian two-state noise are: $\Delta=0.5$,
$\alpha=0.5$, $\langle\tau\rangle=0.01$, $\tau_d=1$. (b) The same data
presented on  log-linear scales.}
\label{Fig3}
\end{figure}

The exact result for the Laplace-transformed relaxation in
Eq. (\ref{eq:solution})  can be reliably inverted numerically
due a generalization of the well-known Stehfest method \cite{Stehfest} in Ref.
\cite{Valko} which requires implementing this method, for example,
with a computer algebra system like
MAPLE (done here) using a sufficiently high digital precision.
As a ``rule of thumb'' the number of digits $N$
used in our calculations
should correspond to the
number of terms taken in the Stehfest asymptotical series expansion \cite{Stehfest}.
$N$ must be increased until the numerical results converge with the required
accuracy. For example, to obtain numerical data  for
Fig. \ref{Fig3} we used $N=256$. In this figure,
the results are numerically
precise within the corresponding line width. Note that the standard
choice $N=16$ \cite{Stehfest} is
inadequate to obtain the correct numerical results for the averaged
relaxation in Fig. {\ref{Fig3}}.
Following the reasoning in \cite{Valko} we checked and confirmed  these numerical
considerations for some test functions with known results for the
``function'' -- and its known Laplace-transform''. Such an 
improved Stehfest method presents one of the best 
numerical Laplace transform inversion methods available nowadays
(the core of the MAPLE code contains just a few lines, by the way).

\begin{figure}
\begin{center}
\includegraphics[width=12cm]{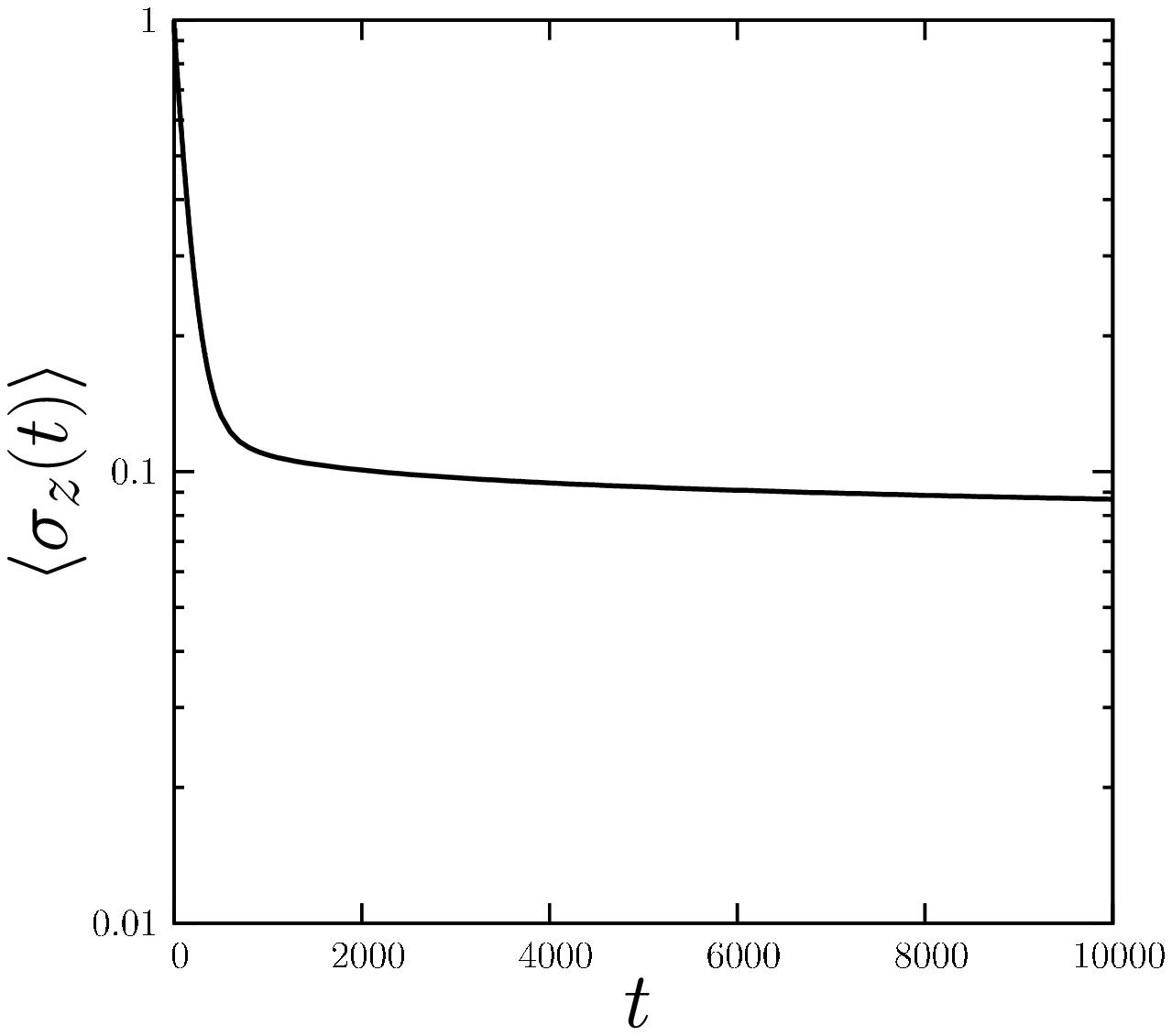}

\vspace{1cm}
\end{center}
\caption{Quantum relaxation under the influence of
non-Markovian two-state noise with a
$1/\omega^{0.9}$ feature for its power spectrum. $\alpha=0.1$, the remaining
parameters are the same as in Fig. \ref{Fig3}. }
\label{Fig4}
\end{figure}

For the results in Fig. \ref{Fig3} the following parameters have been used used: $\epsilon_0$ and
$\Delta$ are the same as in Figs. \ref{Fig1}, \ref{Fig2}, i.e.
$\epsilon_0=1$ and $\Delta=0.5$; $\tau=0.01$
(like for the Markovian case in Fig. \ref{Fig1}), $\alpha=0.5$
(``$1/\omega^{0.5}$'' noise) and $\tau_d=1$.  It is interesting to compare
this case with those in Fig. \ref{Fig1} and Fig. \ref{Fig2}.
Surprisingly, the relaxation dynamics turns out to be initially 
practically a
single exponential with small-amplitude quantum coherent oscillations
superimposed. About 90\% of the initial 
population difference decays exponentially.
 The long time tail of the relaxation process is, however, clearly non-exponential, and the
approach to the steady state occurs much slowly than the initial exponential
decay. The emergence of such a slow non-exponential
asymptotic decay is rather intriguing. It is due to a manifest
non-Markovian character of the noise as detailed  analytically above.
The same reasoning holds true is valid for $\alpha=0.1$;
the corresponding noise obeys an $1/\omega^{0.9}$ feature in its power spectrum which is close
 to  $1/f$ noise. In this case,
the tail of relaxation curve becomes, however, more flat, being in accordance with
Eq. (\ref{asymptotics}), and the
approach to the steady state $\langle \sigma_z(\infty)\rangle = 0$
occurs extremely slow: This might create an  incorrect
impression that $\langle \sigma_z(\infty)\rangle \neq 0$, see
in Fig. \ref{Fig4}.
We remark that for this particular case the digital precision $N=32$
was sufficient to obtain convergent results within the improved Stehfest method.

\section{Resume}

With this  work we have presented general results for the averaged
quantum relaxation that is driven by
discrete state non-Markovian noise of the renewal,
or CTRW type. The noise sources  present  non-Markovian generalization of
 discrete state Markovian noise sources.
Our focus has been on the averaged time-evolution 
in presence of  stationary noise
realizations (i.e. the noise does not relax to, but {\it is} in its
steady state while the noise-averaged quantum dynamics undergoes a
relaxation process).  The practical feasibility of our approach has
been elucidated with several applications of  general interest, namely (i)
the averaging of the Kubo oscillator (for arbitrary noises) and (ii) the averaged relaxation dynamics
of a quantum two state system that is driven by two state non-Markovian noise.
For this latter case and for a symmetric process, tractable analytical
expressions have been obtained in Eqs. (\ref{eq:solution}),
(\ref{eq:solution2}).  The
previously known results for the case of Markovian noise are recovered
from our more general expressions. These new
analytical  results have been verified by 
corresponding numerical studies. In particular, the case of a manifest
non-Markovian noise possessing an infinite variance of the residence time
distribution, an infinite mean autocorrelation time and an $1/\omega^{\alpha}$
feature in its power spectrum has been studied. This latter situation
cannot be tackled within  perturbation theory. However,
our theory allows for a definite non-perturbative treatment. The authors share the confident belief that
this new theory will prove useful for this and similar future investigations
of quantum relaxation processes that are exposed to one or several noisy environments
exhibiting  characteristic  long-time memory.

\section{Acknowledgements}

This work has been  supported in parts by the Deutsche Forschungsgemeinschaft
within the collaborative research centre, SFB-486: ``Manipulation of matter on the nanoscale'', project
A10.


\begin{thebibliography}{100}
\bibitem{Marcus}R.A. Marcus, J. Chem. Phys. 24 (1956) 966.

\bibitem{Hush}N.S. Hush, J. Chem. Phys. 28 (1958) 962.

\bibitem{Levich}V.G. Levich and R.R. Dogonadze, Dokl. Acad. Nauk SSSR,
Ser. Fiz. Khim. 124 (1959) 123 (in Russian).

\bibitem{Hanggi}P. H\"anggi, 
P. Talkner and M. Borkovec, Rev. Mod. Phys. 62 (1990) 251.

\bibitem{Grifoni}M. Grifoni and P. H\"{a}nggi, Phys. Rep. 304 (1998) 229.

\bibitem{Kohler} S. Kohler, J. Lehmann and P. H\"anggi, 
Phys. Rep. 406 (2005) 379.


\bibitem{Jortner}J. Jortner and M. Bixon, eds., Electron Transfer--from
Isolated Molecules to Biomolecules,
Adv.  Chem. Phys. 107, John Wiley and Sons Inc., New York,
1999.

\bibitem{May}
V. May and O. K\"uhn, Charge and Energy Transfer Dynamics in
Molecular Systems, 2nd ed., Wiley--VCH, Berlin, 2004.


\bibitem{Petrov}E.G. Petrov, Physics of Charge Transfer in Biosystems,
Kiev, Naukova Dumka, 1984 (in Russian).


\bibitem{review}I. Goychuk and P. H\"anggi, Adv. Phys. 54 (2005), 525.

\bibitem{Anderson53}P.W. Anderson and P.R. Weiss, Rev. Mod. Phys.
25 (1953) 269.
\bibitem{Kubo54}R. Kubo, J. Phys. Soc. Jpn. 9 (1954) 935.

\bibitem{Kubo62}
R. Kubo, in: Fluctuation, Relaxation, and Resonance in Magnetic
Systems, ed. by D. ter Haar, Oliver and Boyd, Edinburgh, 1962.

\bibitem{Kubo63}R. Kubo, J. Math. Phys. 4 (1963) 174.

\bibitem{Kubo69}R. Kubo, Adv. Chem. Phys. 15 (1969) 101.

\bibitem{Burshtein}A.I. Burshtein, Zh. Eksp. Teor. Phys. 49
(1965) 1362 [Sov. Phys. JETP 22 (1966) 939].

\bibitem{Haken72}H. Haken and P. Reineker,
Z. Phys. 249 (1972) 253.

\bibitem{Haken73} H. Haken and G. Strobl,
Z. Phys. 262 (1973) 135.

\bibitem{BlumenSilbey}A. Blumen and R. Silbey, J. Chem. Phys. 69 (1978)
3589.

\bibitem{Fox} R. F. Fox, Phys. Rep. 48 (1978) 181.

\bibitem{MukamelOppenRoss}S. Mukamel, I. Oppenheim, and J. Ross,
Phys. Rev. A 17 (1978) 1988.

\bibitem{ReinekerBook}P. Reineker,  Stochastic Liouville Equation
Approach -
Coupled Coherent and Incoherent Motion,
Optical Line Shapes, Magnetic-Resonance Phenomena, in:
V. Kenkre and P. Reineker, Exciton Dynamics in Molecular Crystals
and Aggregates,
Springers Tracts in Modern Physics, vol. 94, Springer, Berlin, 1982.

\bibitem{LindWest}K. Lindenberg and B. West, The Nonequilibrium
Statistical Mechanics of Open and Closed Systems, VCH, New York, 1990.


\bibitem{Stockburger} J. T. Stockburger and H. Grabert, 
Chem. Phys. 268 (2001) 249.

\bibitem{Shao}J. Shao, J. Chem. Phys. 120 (2004) 5053.


\bibitem{OvchinErikh}A.A. Ovchinnikov and N.S. Erikhman,
Zh. Exsp. Teor. Phys.  67 (1974) 1474 (in Russian)
[Sov. Phys. JETP  40 (1975) 733].

\bibitem{Kayanuma84}Y. Kayanuma, J. Phys. Soc. Jpn.  53  (1984) 108.
1984
\bibitem{Kayanuma85}Y. Kayanuma, J. Phys. Soc. Jpn. 54 (1985) 2037.

\bibitem{ShaoZerbeHan98}J. Shao, C. Zerbe, and P. H\"anggi,
Chem. Phys.  235 (1998) 81.

\bibitem{HanggiJung}P. H\"anggi and P. Jung, Adv. Chem. Phys. 89 (1995)
239.

\bibitem{KampenBook}N.G. Van Kampen,
Stochastic Processes in Physics and
Chemistry, 2d ed., North-Holland, Amsterdam, 1992.


\bibitem{Schulten}R. Bittl and K. Schulten, J. Chem. Phys. 90 (1989)
1794.


\bibitem{FrischBrissaud1} U. Frisch and A. Brissaud,
J. Quant. Spectrosc. Radiat. Transfer  11 (1971) 1753.

\bibitem{Wiener}N. Wiener, Ann. Math. 22 (1920) 66.


\bibitem{FeynmannBook}R.P. Feynman and A.R. Hibbs, Quantum Mechanics
and Path Integrals, McGraw-Hill Book Company, New York, 1965.

\bibitem{FrischBrissaud2}
A. Brissaud and U. Frisch, J. Math. Phys. 15 (1974) 524.

\bibitem{Hanggi78}
P. H\"anggi, Z. Physik B 31 (1978) 407.

\bibitem{Hanggi80}
P. Hanggi, Z. Physik B 36 (1980) 271.

\bibitem{Hanggi81} P. Hanggi, Z. Physik B 43 (1981) 269.

\bibitem{ShapLog1} V.E. Shapiro and V.M. Loginov, Dynamical
Systems under Stochastic Perturbations: Simple Methods of Analysis, 
Nauka, Novosibirsk, 1983 (in Russian).

\bibitem{Klyatskin}
V.I. Klyatskin, Radiophys. Quant. Electron.  20 (1978) 382
[Izv. Vyssh. Ucheb. Zaved., Radiophyz. 20 (1977) 562].

\bibitem{ShapLog}V.E. Shapiro and V.M. Loginov, Physica A 91 (1978)
563.

\bibitem{HanggiProceedings} P. H\"anggi, in:
Stochastic Processes Applied to Physics, ed. by L. Pesquera
and M.A. Rodriguez, World Scientific, Singapore, 1985; pp. 69 - 95.

\bibitem{AverPouq}
M. Auvergne and A. Pouquet, Physica 66 (1973) 409.


\bibitem{Reineker1}V. Kraus and P. Reineker, Phys. Rev A 43 (1991)
4182.

\bibitem{Reineker2}P. Reineker, Ch. Warns, Th. Neidlinger, and I. Barvik,
Chem. Phys. 177 (1993) 713.

\bibitem{PRE95_3} I.A. Goychuk, E. G. Petrov and V. May,
Phys. Rev. E 51 (1995) 2982.

\bibitem{PLA96} E. G. Petrov, I.A. Goychuk, V. Teslenko and V. May,
Phys. Lett. A 218 (1996) 343.

\bibitem{Iwan}J. Iwaniszewski, Phys. Rev. E 61 (2000) 4890.

\bibitem{AnkerPech}J. Ankerhold and P. Pechukas,
Europhys. Lett. 52 (2000) 264.

\bibitem{Reineker3}I. Barvik, C. Warns, T. Neidlinger, and P. Reineker,
Chem. Phys. {\bf 240} (1999) 173.

\bibitem{JCP95}I.A. Goychuk, E.G. Petrov, and V. May,
J. Chem. Phys. 103 (1995) 4937.


\bibitem{Kampen79}N.G. van Kampen, Physica A 96 (1979)  435.

\bibitem{Bursh86}A. I. Burshtein, A. A. Zharikov, S. I. Temkin,
Theor. Math. Phys. {\bf 66}, 166 (1986).

\bibitem{Chvosta}P. Chvosta and P. Reineker, Physica A 268 (1999)
103.


\bibitem{Feller}W. Feller, An Introduction to Probability Theory
and its Applications, 2nd corrected ed., Volume II,
John Wiley, New York, 1966.

\bibitem{Goychuk04}I. Goychuk, Phys. Rev. E 70 (2004) 016109.

\bibitem{MontrolWeiss}E.W. Montroll and G. H. Weiss, J. Math.
Phys. 6 (1965) 167.


\bibitem{LaxSher}M. Lax and H. Scher,  Phys. Rev. Lett. 39 (1977)
781.

\bibitem{Shlesinger}M.F. Shlesinger,
in: Encyclopedia of Applied Physics 16, VCH, New York, 1996, pp. 45-70.

\bibitem{Hughes}B.D. Hughes,  Random Walks and Random Environments,
Volume 1, Clarendon Press, Oxford, 1995.

\bibitem{cox}D.R. Cox, Renewal Theory, Methuen, London, 1962.

\bibitem{JungSilbey}Y. Jung, E. Barkai, and R. J. Silbey, Adv. Chem. Phys.
123 (2002) 199.

\bibitem{JCP05}I. Goychuk, J. Chem. Phys. 122 (2005) 164506.

\bibitem{JungBarkSilb} Y. Jung, E. Barkai, and R. J. Silbey, Chem. Phys.
284 (2002) 181.

\bibitem{PRE94} E.G. Petrov, V.I. Teslenko, and I.A. Goychuk,
 Phys. Rev. E 49 (1994) 3894.

\bibitem{PRL98}I. Goychuk, M. Grifoni, and P. H\"anggi,
Phys. Rev. Lett. 81 (1998) 649;
Phys. Rev. Lett.  81 (1998) 2837 (erratum).

\bibitem{ReilSkin}P.D. Reilly and J.L. Skinner, J. Chem. Phys. 101 (1994)
959.

\bibitem{GevaSkinner}E. Geva and J.L. Skinner, Chem. Phys. Lett.
288 (1998) 225.

\bibitem{Barkai}E. Barkai, R. Silbey, and G. Zumofen,
Phys. Rev. Lett. 84 (2000) 5339.

\bibitem{Strat} R. L. Stratonovich, Topics in the Theory of
Random Noise, Volume I, Gordon and Breach, New York, 1963, p. 176.

\bibitem{LowenTeich}S.B. Lowen and M.C. Teich, Phys. Rev. E
47 (1993) 992.

\bibitem{PRL03} I. Goychuk and P. H\"anggi, Phys. Rev. Lett.
91 (2003) 070601.

\bibitem{PRE04}  I. Goychuk and P. H\"anggi, Phys. Rev.
E 69 (2004) 021104.

\bibitem{Wilhelm}
 H. Gutmann, F.K. Wilhelm, W.M. Kaminsky, and S. Loyd,
 Phys. Rev. A 71 (2005) 020302.

\bibitem{Slichter}C. P. Slichter, Principles of Magnetic Resonance,
Springer, Berlin, 1978.



\bibitem{Stehfest}H. Stehfest,
Comm. ACM 13 (1970) 47; Comm. ACM 13 (1970) 624 (Erratum).

\bibitem{Valko}
P. P. Valko and S. Vajda,
Inverse Problems in Engineering 10 (2002) 467.

\bibitem{Metzler}R. Metzler, J. Klafter, J. Jortner,  Proc. Natl. Acad.
Sci. USA 96 (1999) 11085.

\bibitem{Sansom}M.S.P. Sansom, F.G. Ball, C.J. Kerry, R. McGee,
R.L. Ramsey, and P.N.R. Usherwood, Biophys. J. 56 (1989) 1229.


\bibitem{PREfrac}I. Goychuk and P. H\"anggi, Phys. Rev. E 70 (2004)
051915.

\end{thebibliography}
\end{document}